 \documentclass{elsart}

\usepackage{graphicx}
\usepackage{amssymb}

\usepackage[english,francais]{babel}

\newtheorem{e-proposition}[theorem]{Proposition}

\newtheorem{e-definition}[theorem]{Definition\rm}

\newcommand{\ltappr}{{{\lower4pthbox{$<$} } \atop \widetilde{ \ \ \ }}}
\newlength{\bxwidth}\bxwidth=1.5 truein

\newcommand{\dg}{^{\dagger }}

\newcommand{\rarrow}{\rightarrow}
\newcommand \bea {\begin{eqnarray} }
\newcommand \eea {\end{eqnarray}}
\newcommand{\bk}{{\bf{k}}}

\newcommand{\down}{{j\downarrow}}

\newlength{\figwidth}
\newlength{\shift}
\newcommand{\myfg}[3]
{
\begin{figure}[ht]

\vspace*{-0cm}
\[
\includegraphics[width=\figwidth]{#1}
\]
\vskip -0.2cm
\caption{\label{#2}
\small#3
}
\end{figure}}

\def\og{\leavevmode\raise.3ex\hbox{$\scriptscriptstyle\langle\!\langle$~}}
\def\fg{\leavevmode\raise.3ex\hbox{~$\!\scriptscriptstyle\,\rangle\!\rangle$}}

\begin{document}
% Select a primary header Physics or Astrophysics
% You can place after the header (classification), if you know it.

%\centerline{Physics Header}
%\begin{frontmatter}


% \author{Name\thanksref{label2}}
\selectlanguage{english}

\title{Spins, electrons and broken symmetries: realizations of two channel Kondo physics}

% If all authors are at the same address, the [label1] can be suppressed

\selectlanguage{english}
\author[authorlabel1]{Rebecca Flint}
\ead{flint@iastate.edu}
\author[authorlabel2,authorlabel3]{Piers Coleman}
\ead{coleman@physics.rutgers.edu}

\address[authorlabel1]{Department of Physics and Astronomy, Iowa State University, 12 Physics Hall, Ames, Iowa 50011
USA}
\address[authorlabel2]{Center for Materials Theory, Department of Physics and Astronomy, Rutgers University, 136
Frelinghuysen Rd, Piscataway, New Jersey 08854 USA}
\address[authorlabel3]
{Department of Physics, Royal Holloway, University of London, Egham, Surrey, TW20 0EX, UK}

% If your know the dates of reception, and acceptation you can put them now;
%    idem the name of the person presenting your article

%\medskip
%\begin{center}
%{\small Received *****; accepted after revision +++++}
%\end{center}

\begin{abstract}

Adding a second Kondo channel to heavy fermion materials reveals new exotic symmetry breaking phases associated with the development of Kondo coherence.  In this paper, we review two such phases, the ``hastatic order'' associated with non-Kramers doublet ground states, where the two-channel nature of the Kondo coupling is guaranteed by virtual valence fluctuations to an excited Kramers doublet, and ``composite pair superconductivity,'' where the two channels differ by charge $2e$ and can be thought of as virtual valence fluctuations to a pseudo-isospin doublet.  The similarities and differences between these two orders will be discussed, along with possible realizations in actinide and rare earth materials like URu$_2$Si$_2$ and NpPd$_5$Al$_2$.}
%{\it To cite this article: R. Flint, P. Coleman, C. R. Physique 6 (2005).}

\vskip 0.5\baselineskip

%Now keywords/mots-cl�s
%\noindent{\small{\it Mots-cl\'es~:} Mot-cl\'e1~; Mot-cl\'e2~;Mot-cl\'e3}}
\end{abstract}
%\end{frontmatter}
\maketitle

\selectlanguage{english}
% main text
\section{Introduction}
\label{}

The interplay of nearly free conduction electrons and localized f-electrons in heavy fermion materials gives rise to a fascinating competition between magnetism and the heavy Fermi liquid resulting from the hybridization of c- and f-electrons\cite{mott}.  This competition is thought to generate rich phase diagrams containing not only heavy Fermi liquids and magnetically ordered phases\cite{doniach}, but superconductivity and exotic spin liquid phases\cite{mathur08,extended}.  All of this physics emerges from a single Kondo channel - a single symmetry in which conduction electrons can screen the local moments.  When a second Kondo channel is added, the physics 
is potentially even richer.  This new physics is particularly relevant for actinide materials, where the larger 5f
orbitals lead to more mixed valency than their rare-earth cousins, with correspondingly
higher temperature scales. In this paper we review
two exotic new phases proposed to result from the interference of competing
screening channels: ``hastatic order'' associated with the two screening
channels of a non-Kramers doublet and ``composite pairing'' occuring when a
Kramers double interacts with two different channels, one hole-like ($c$) and
the other electron-like ($c\dg$).  Both proposals are motivated by real materials: hastatic order is a possible explanation of the hidden order in URu$_2$Si$_2$\cite{palstra85} and composite pair superconductivity may explain how superconductivity can arise directly out of a Curie paramagnet in certain ``115'' materials like Ce$M$In$_5$ ($M$ = Co,Ir)\cite{petrovic,cecoir} and NpPd$_5$Al$_2$\cite{aoki08}.

\myfg{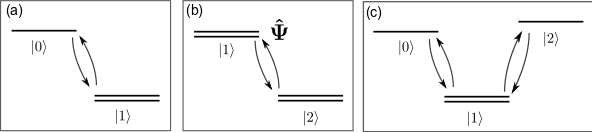}{kondos}{(a) The usual Kondo effect involves virtual valence fluctuations between a Kramers doublet and an excited singlet state. (b) Hastatic order involves a non-Kramers doublet fluctuating to an excited Kramers doublet. (c) Composite pairing arises when a Kramers doublet fluctuates to two excited singlets whose charge differs by $2e$.}

Heavy fermion materials contain two species of electrons: nearly free conduction electrons and strongly interacting f-electrons that are localized at high temperatures. The Kondo effect is an antiferromagnetic interaction through which the conduction electrons screen the local moments to form Kondo singlets, giving rise to a heavy Fermi liquid.  The Kondo effect can also be thought of as a hybridization between two types of electrons; however, as the f-electrons are strongly interacting, the object hybridizing with the conduction electrons is not the original f-electron, but rather a composite fermion consisting of a conduction electron and a spin flip, $f_{\uparrow}\dg \sim c_{\downarrow}\dg S_+$.  In the single-channel Kondo effect, this hybridization is generated by valence fluctuations of the f-ion from a ground state doublet to an excited singlet state, as shown in Fig 1(a). As the excited singlet carries no quantum numbers, it breaks no symmetries and the Kondo effect develops as a crossover. This process is captured in the single-channel Anderson model,
\begin{equation}
H = \sum_\bk \epsilon_\bk c\dg_\bk c_\bk + V \sum_j \left(c_j\dg|0\rangle\langle \sigma| + H.c.\right) + \sum_j \epsilon_f |\sigma\rangle\langle \sigma|
\end{equation}
where $|0\rangle$ and $|\sigma \rangle$ represent the empty (excited) and singly-occupied (ground) states of the f-ion, and the doubly occupied states, $|2\rangle$ are forbidden.  Typically, we solve this model by introducing a slave boson, $b\dg|\Omega\rangle$ to represent the excited singlet, $|0\rangle$ and a pseudo-fermion, $f\dg_\sigma|\Omega\rangle$ to represent the ground state doublet, $|\sigma\rangle$, where $|\Omega\rangle$ is the particle vacuum\cite{coleman83}. The development of a coherent Kondo effect is then captured by the development of $\langle b \rangle$ at the Kondo temperature, $T_K$, which decreases the valence, $n_f = 1-\langle b \rangle^2$. In this mean field approach, the Kondo affect appears as a phase transition, but as it is not protected by symmetry, gauge fluctuations restore it to a crossover\cite{newns}.

While the usual Kondo effect involves an excited singlet, the two-channel Kondo effect involves an excited doublet, protected by channel symmetry.  The development of Kondo coherence breaks this channel symmetry, causing the coherence to onset at a phase transition rather than a crossover.  Typically the channel symmetry will coincide with another physical symmetry; in our two examples, these are time-reversal and particle-hole symmetry.  These two symmetries describe the two main classes of two-channel Kondo problems, and can be distinguished by the number of f-electrons.  The single-channel Kondo effect typically results from materials with an odd number of f-electrons, where the ground state is guaranteed to be a Kramers doublet, protected by time-reversal symmetry.  The excited states contain even numbers of f-electrons and are usually taken to be singlets, unprotected by time-reversal symmetry.  

Atoms with even numbers of f-electrons can also have doublet ground states; these non-Kramers doublets are protected by crystal symmetry rather than time-reversal symmetry.  Here, valence fluctuations involve excited states with an odd number of f-electrons: Kramers doublets.  This scenario is illustrated in Fig 1(b), where now the two excited states each require a slave boson, $\hat \psi_\uparrow$ and $\hat \psi_\downarrow$ that can be packaged into a spinor, $\hat \Psi = (\hat \psi_\uparrow, \hat \psi_\downarrow)$.  The development of Kondo coherence, $\langle \hat \Psi\rangle$ requires the spinor to pick a direction in spin-space, breaking both spin-rotation and time-reversal symmetries.  In fact, as here the ``order parameter'' $\hat \Psi$ carries a half-integer spin and behaves as a spinor, the resulting state is more subtle than the conventional, vectorial magnetic order.  This spinorial hybridization is what we have termed hastatic order\cite{chandra13}; the primary order parameter is the hybridization gap, but all other symmetry breaking observables are suppressed by $T_{HO}/D$, where hastatic order onsets at $T_{HO}$ and $D$ is the bandwidth, making this state difficult to observe, and a good candidate to explain hidden order in URu$_2$Si$_2$.

The ground state Kramers doublet can also exhibit two channel Kondo physics; one case involves valence fluctuations to two excited singlets that differ by charge $2e$: $f^{n-1} \leftrightarrow f^n \leftrightarrow f^{n+1}$, as shown in Fig 1(c).  If these two excited singlets have the same energy, they form an isospin doublet. The two channel Kondo effect then breaks U(1) charge conjugation symmetry to form a composite pair superconductor\cite{catk,flint08}, where the Kondo temperature becomes the superconducting transition temperature, $T_c$.  Composite pairs are to Cooper pairs what composite fermions are to electrons - they incorporate a local moment spin-flip, $\Delta_C \sim \langle c\dg_{1\uparrow} c\dg_{2\uparrow} S_-\rangle$, and here two conduction electrons in different channels (1,2) screen the \emph{same} local moment, creating a local pair\cite{abrahams}.  

\figwidth=8.5cm
\myfg{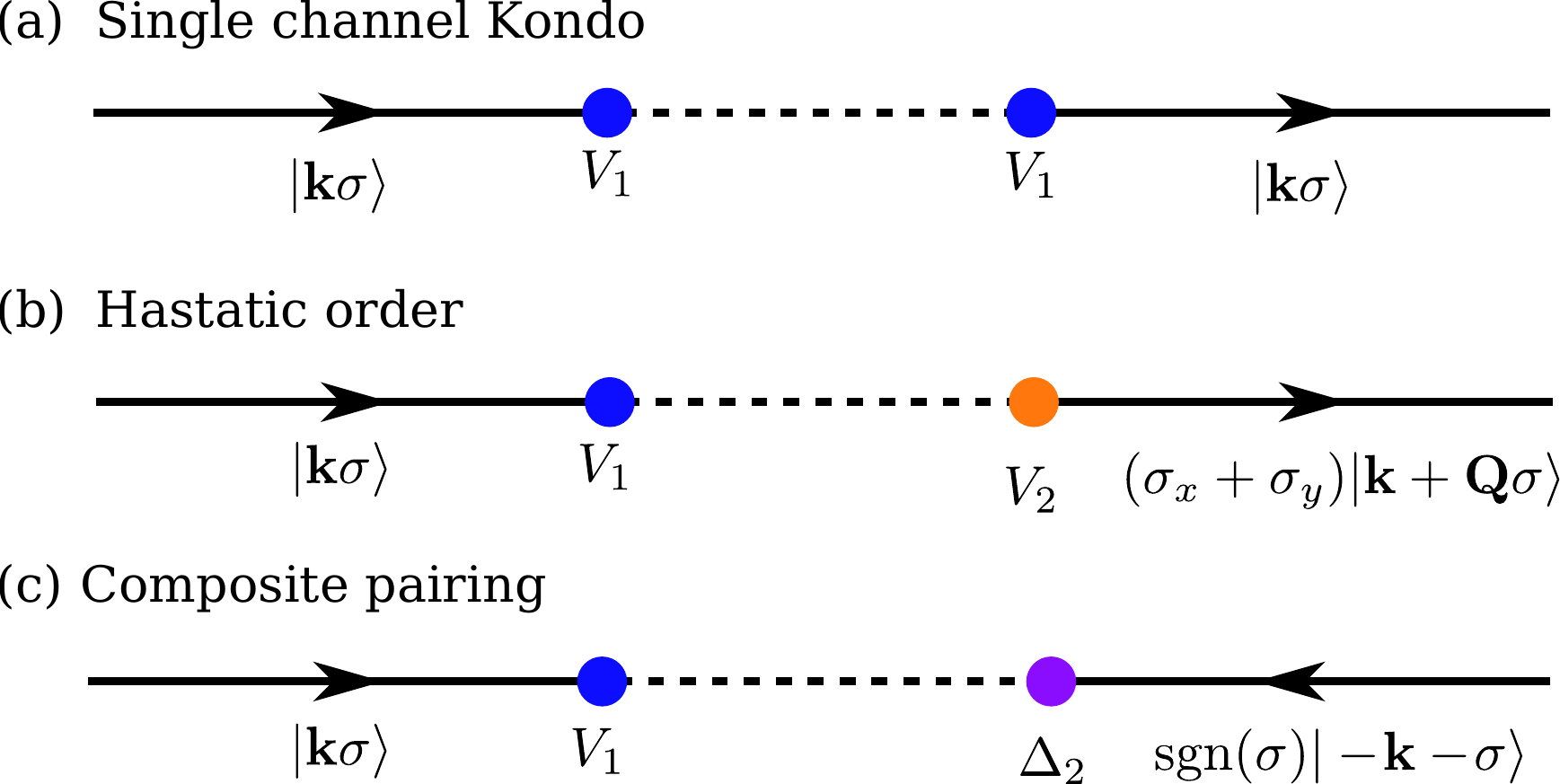}{scattering}{The two channel Kondo physics captured in the Anderson models shown in Fig 1 can also be treated in a Kondo picture, where there are two channels for scattering a conduction electron (solid line) off a single f-ion (dashed line).  (a) Shows the usual single channel Kondo scattering, that conserves momentum and spin. $V_1$ represents the effective hybridization between $c$ and $f$, $V_1 \sim \langle c_1\dg f\rangle$, where $c_1\dg$ creates a conduction electron with symmetry $\Gamma_1$.  When we introduce a second channel with symmetry $\Gamma_2$, we have two types of hybridization: electron-hole, $V_2 \sim \langle c_2\dg f\rangle$ and electron-electron $\Delta_2 \sim \langle c_2 f\rangle$.  While intra-channel scattering cannot break symmetries, inter-channel scattering can. (b) Hastatic Kondo scattering, where the scattering breaks time-reversal, multiplying the original conduction electron by a linear combination of $\sigma_x$ and $\sigma_y$, and translation symmetries, adding a momentum ${\bf Q}$. (c) Composite pair Kondo scattering, where the electron Andreev scatters off the Kondo impurity, a process requiring broken $U(1)$ gauge symmetry and the presence of a condensate of composite pairs.}

While the distinct broken symmetries mean the two phases appear quite different, their Kondo origins lead to several key similarities. Above $T_K \equiv T_{HO}, T_c$, the local moments are mainly unquenched, leading to a Curie-Weiss susceptibility that is quenched at $T_{HO}$ or $T_c$ and a large entropy of condensation related to the $\frac{1}{2} R \log 2$ zero point entropy of the two channel Kondo impurity. Real systems will include fluctuations missing from the mean-field calculations that will partially quench the moments above the transition temperature, possibly quite differently for the two phases. Both phases involve the development of a hybridization gap - for hastatic order, this is the typical hybridization gap centered either above or below the Fermi energy, $E_F$, while for composite pairing, the superconducting gap is a hybridization gap, pinned at $E_F$.  And both phases will be suppressed in magnetic field, as is the usual Kondo effect; indeed,  CeCoIn$_5$\cite{ceQCP}, NpPd$_5$Al$_2$\cite{npQCP} and URu$_2$Si$_2$\cite{uQCP} all share a quantum critical point at the critical field, whose non-Fermi liquid behaviors may be a remnant of the original two-channel Kondo critical fluctuations.

\section{Non-Kramers ground state: Hastatic order}

The problem of hidden order in URu$_2$Si$_2$ is one of the oldest in condensed matter.  At high temperatures, URu$_2$Si$_2$ looks like a typical heavy fermion material with Ising magnetic moments. However, at $T_{HO} = 17.5K$, it undergoes a mean-field-like phase transition involving nearly one-third of the spin entropy\cite{palstra85}. The order parameter developing at this phase transition has eluded identification for over 27 years, leading to the name ``hidden order''\cite{mydosh11}.

While there is currently no consensus on the relative importance of itinerant and local physics in URu$_2$Si$_2$, or even on the dominant valence of the uranium ion, with various probes suggesting either 5f$^2$ or 5f$^3$\cite{broholm91,park02,jeffries10}, the large magnetic anisotropy seen both in the high temperature susceptibility\cite{palstra85} and in de Haas-van Alphen (dHvA) measurements at low temperatures\cite{altarawneh12} is difficult to reconcile with a Kramers doublet ground state. In particular, the observation of Ising-like \emph{conduction electrons} at low temperatures, suggests that the conduction electrons must be hybridized with an Ising, and thus non-Kramers doublet.  The possible non-Kramers doublet\cite{amitsuka94},
\begin{equation}
|\Gamma_5 \pm \rangle = a|J = 4, J_z = \pm 3\rangle + b |J = 4, J_z =\mp 1\rangle
\end{equation}
is always Ising-like, while 5f$^3$ Kramers doublets are Ising-like only when finely-tuned\cite{flint13}. Therefore, we believe the URu$_2$Si$_2$ ground state to be a non-Kramers doublet, and its hybridization must therefore break time-reversal symmetry.

There are several recent experiments hinting that the hidden order involves hybridization: STM experiments find that the hybridization gap and the heavy band development at $T_{HO}$\cite{schmidt10,aynajian10}; pump-probe optical measurements that find the quasiparticle lifetime increasing sharply below $T_{HO}$\cite{dakovski11}; and dHvA finds that the heavy quasiparticles at low temperatures have a strong Ising anisotropy inherited from the f-electrons\cite{altarawneh12}. These results indicate that the hidden order is a hybridization between Ising (non-Kramers) f-electrons and (inherently Kramers-like) conduction electrons. The most generic valence fluctuation term capturing this kind of hybridization is,
\begin{equation}
H = V_{\sigma \alpha} |\bk \sigma\rangle \langle \alpha| + H.c.
\end{equation}
where $|\bk\sigma\rangle$ represents a conduction electron with spin $\sigma$ and momentum $\bk$, $|\alpha\rangle$ represents a $\Gamma_5$ state with pseudo-spin $\alpha$ and $V_{\sigma \alpha}$ the hybridization between them. The key difference between Kramers and non-Kramers states is their behavior under double-time-reversal symmetry.  Kramers states pick up a negative sign, $|\bk\sigma\rangle \stackrel{\theta^2}{\longrightarrow} -|\bk\sigma\rangle$, while non-Kramers states are left invariant,
$|\alpha\rangle \stackrel{\theta^2}{\longrightarrow} +|\alpha\rangle$.  Since the Hamiltonian is trivially invariant under double-time-reversal, $V_{\sigma \alpha}$ must invert under double time-reversal, $V_{\sigma \alpha} \stackrel{\theta^2}{\longrightarrow} -V_{\sigma \alpha}$, and so $V_{\sigma \alpha}$ transforms like a \emph{spinor}, breaking both single and double time-reversal.  In otherwords, $V_{\sigma \alpha}$ mixes a half-integer spin state with an integer spin state and must itself carry a half-integer spin - this is the slave boson spinor representing the excited Kramers doublet.  It is this spinorial hybridization that characterizes hastatic order\cite{chandra13}.

When this hastatic spinor orders, it develops not only a magnitude, the usual Kondo effect, but also selects a direction in spin space, breaking both time-reversal and spin-rotation symmetries.  If the spinor is staggered and points along the magnetic c-axis, the resulting state is an antiferromagnet - actually a hastatic antiferromagnet where the large f-electron magnetic moments develop as a consequence of the hybridization, not magnetic ordering, although this state is mostly indistinguishable from a conventional antiferromagnet. If the hastatic spinor instead points in the basal plane, the resulting state has no large moments, and in fact strongly resembles the hidden order; this state is what we call hastatic order. As hastatic order is related to antiferromagnetism by a rotation, there is a first-order 'spin-flop' transition between the two (Fig. 3), and so the hastatic picture easily captures the pressure phase diagram of URu$_2$Si$_2$\cite{villaume08}.  The longitudinal spin fluctuations of the hastatic spinor vanish with a square root behavior at this first-order transition, $\Delta \sim \sqrt{T-T_c}$, one of the key predictions of hastatic order.  Hastatic order has several other experimental consequences. As the non-Kramers doublet is protected by tetragonal and time-reversal symmetries, the hybridization breaks both.  Broken time-reversal symmetry leads to a staggered basal plane conduction electron moment whose magnitude is limited by $T_{HO}/D$, where $D$ is the conduction electron bandwidth, and has not yet been observed in URu$_2$Si$_2$, although recent neutron experiments suggest that any such moment must be smaller than $0.001\mu_B$\cite{das13,metoki13,ross14}, indicating a very small degree of mixed valency\cite{chandra14}; this magnitude of moment is consistent with NMR and RXS experiments\cite{bernal,caciuffo}.  Broken tetragonal symmetry has already been found, both as the development of a nonzero $\chi_{xy}$\cite{okazaki11} and as a tiny orthorhombic distortion\cite{matsuda}.  The quenching of the Curie-Weiss susceptibility at around 70K suggests that hastatic order melts via phase fluctuations, where the amplitude of the hybridization spinor develops at 70K, but the direction of the spinor remains disordered until the symmetry is broken at $T_{HO}$, as shown in Fig 3.

\figwidth=7cm
\myfg{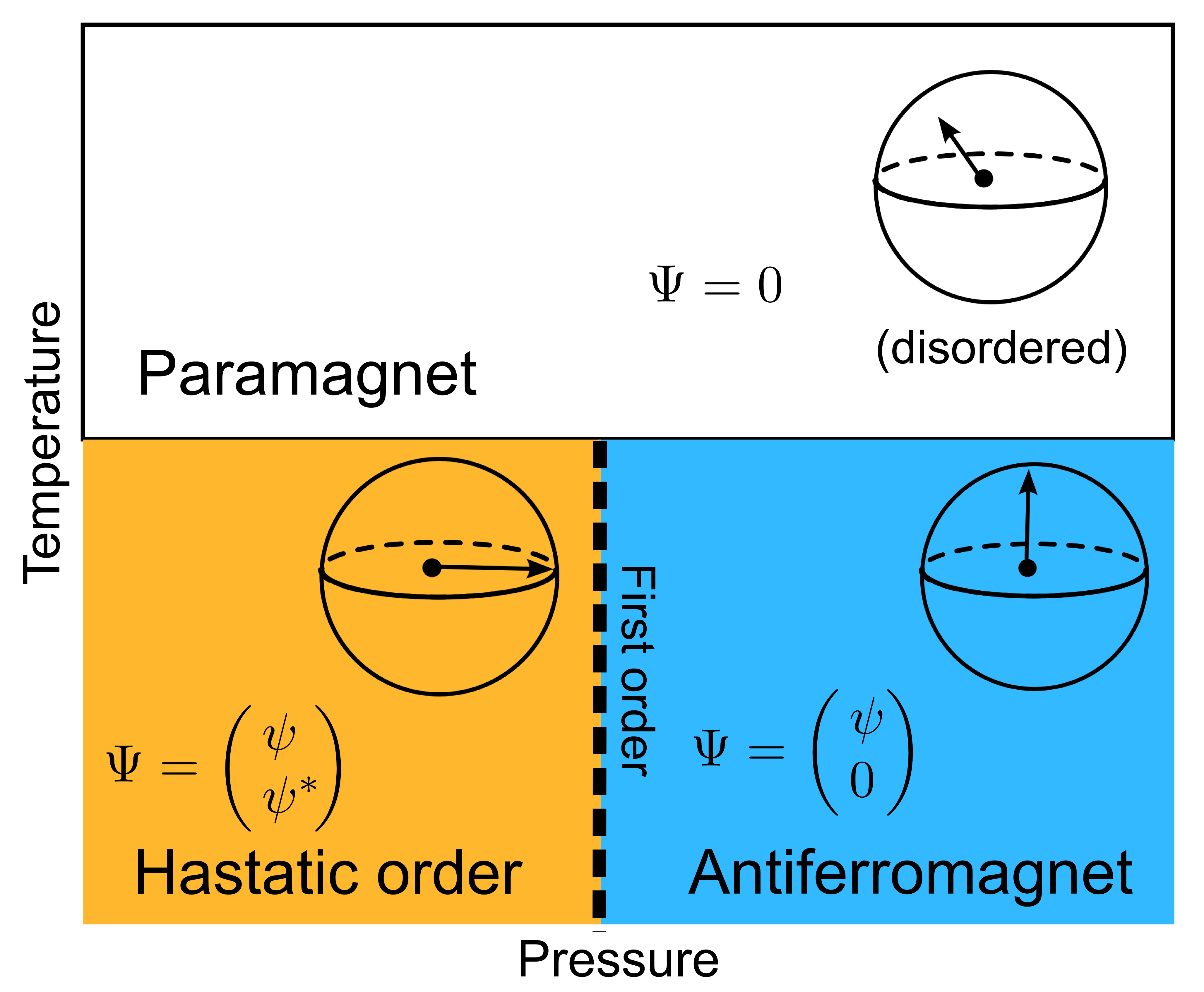}{HO}{Hastatic order phase diagram: the hybridization spinor is disordered in the paramagnet, points in the basal plane for hastatic order and along the c-axis in the antiferromagnet.}

\section{Kramers ground state: Composite pair superconductivity}

Composite pairing is generated by the two-channel Kondo effect involving two different charges\cite{catk,flint08}. The first Kondo effect forms a Kondo resonance as electrons scatter off the local moments; the second channel allows that resonance to itself resonate between electron and hole channels, creating a condensate of pairs; alternately, composite pairing can be thought of as an Andreev scattering of the conduction electrons off of the local moments, as shown in Fig 2 (c).  The pairing is strongest when the two channels have equal strengths, however, as the pairing term shares the Cooper logarithm, the ground state will always be superconducting, although $T_c$ is exponentially suppressed by the ratio of the channel strengths.  The involvement of the local moments means that these are composite pairs, $\langle c_{1\down}\dg c_{2\down}\dg S_+\rangle$, which combine a triplet pair of conduction electrons in two orthogonal symmetries (1 and 2) with a local spin flip to make a charge $2e$ singlet. The particular symmetries of the two channels (determined by crystal fields) determine the symmetry of the pair; for the $J=5/2$ Ce 115s, channel one is $|\Gamma_7^+ \pm\rangle \sim |J_z = \pm 5/2\rangle$, while the second channel is thought to be $|\Gamma_6 \pm\rangle = |J_z = \pm 1/2\rangle$, and so as the conduction electrons Andreev scatter, they must pick up two units of angular momentum, creating a $d_{x^2-y^2}$-like composite pair\cite{flint10}. These singlet, d-wave composite pairs have all the symmetries of a magnetic pair and so the two mechanisms can work in tandem to raise $T_c$, as shown in the phase diagram of Fig. 4(b)\cite{flint10}. The presence of a second mechanism in Ce (where 4f$^1$ fluctuates to both 4f$^0$ and 4f$^2$), but not in Yb (where 4f$^{13}$ only fluctuates to 4f$^{14}$) naturally explains the dearth of Yb superconductors. The presence of a second mechanism also explains how there can be two superconducting domes in Ce$M$In$_5$, as $M$ is tuned from Rh to Ir to Co\cite{sarrao07}, presumably changing the relative ratios of magnetic and composite pairing.

\myfg{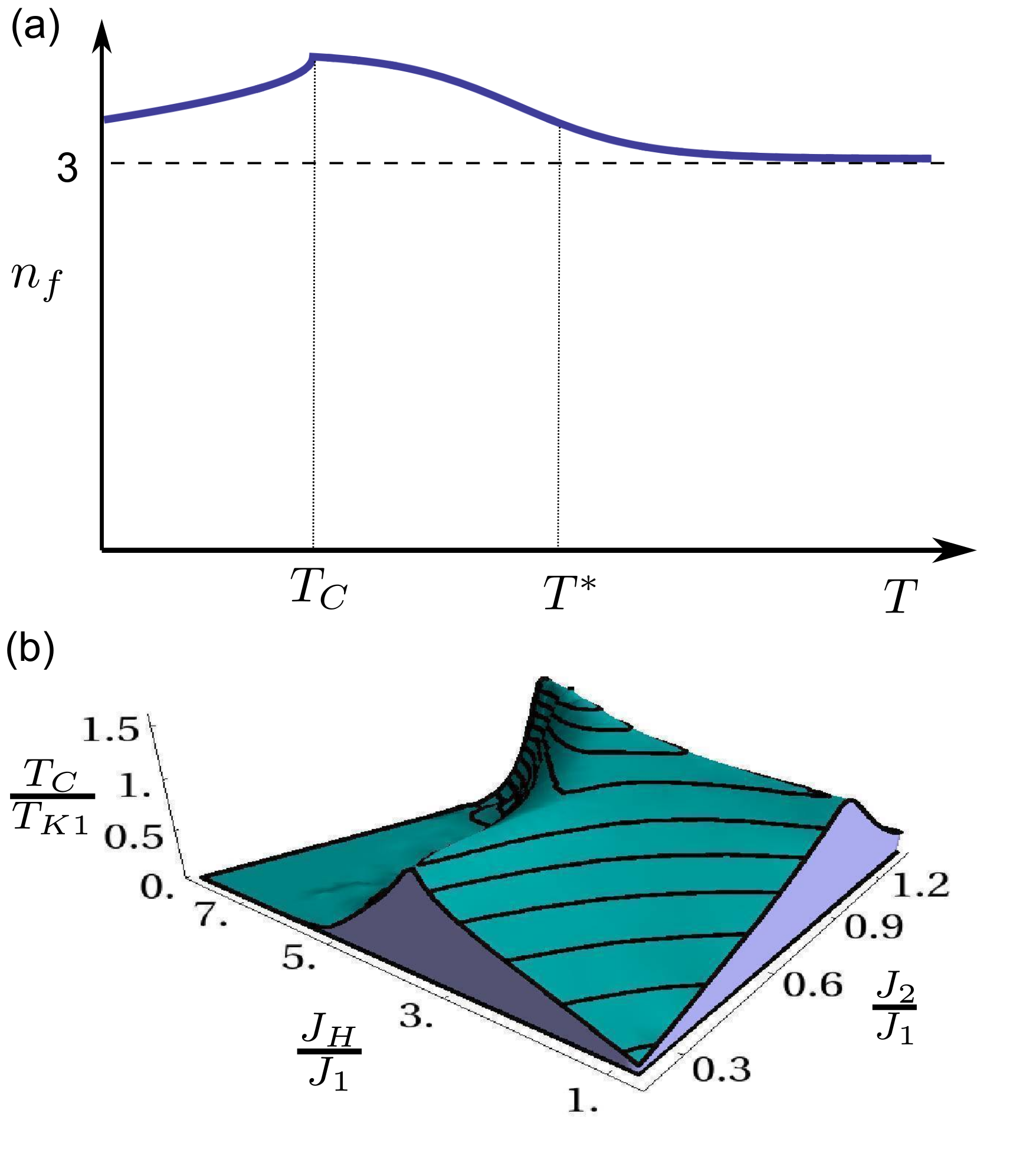}{composite'}{(a) Expected $n_f(T)$ behavior in NpPd$_5$Al$_2$, where there is a smooth crossover at $T^*$, but a kink at $T_c$. (b) The phase diagram for the two channel Kondo-Heisenberg model, that captures both magnetic pairing (favored by the magnetic coupling $J_H$ and composite pairing (strongest when the two Kondo couplings, $J_1$ and $J_2$ are equal), and how they work together to increase $T_c$.}

As composite and magnetic pairs are identical from a symmetry perspective, a key question is how to distinguish them. There are several important differences: first, composite pairing can emerge directly out of a Curie paramagnet, as seen in CeCoIn$_5$\cite{petrovic} and NpPd$_5$Al$_2$\cite{aoki08}, which is difficult to obtain in a magnetic scenario.  Secondly, composite pairing is a local phenomena, taking place mainly within a single unit cell, and as such it should be far more robust to disorder on the rare earth sites (which disturb only a single unit cell) than to disorder on the In sites (which disturb multiple unit cells); indeed, superconductivity persists up to 80\% doping of Yb on the Ce site in CeCoIn$_5$\cite{paglione07}, but is suppressed by 3\% doping of Sn on In\cite{sn}.  Finally, the Kondo nature of composite pairing means that it affects the charge of the f-ion, both by changing the valence and as higher multipole moments of the charge distribution\cite{flint11}. The valence, $n_f(T)$ changes smoothly through $T^*$, but composite pairing develops as a phase transition and so leads to a sharp kink at $T_c$. The f-valence can be measured by core-level x-ray spectroscopy in the Ce 115s and the M\H{o}ssbauer isomer shift in NpPd$_5$Al$_2$\cite{gofryk09}. Similarly, as the electron and hole channels involve  f-electron orbitals with different symmetries, the composite pair condensate carries a quadrupole moment.  This moment also develops sharply at $T_c$ and can be measured with NQR. We estimated the NQR frequency shift to be $\approx 5$ kHz/K\cite{flint11}, and a shift of this magnitude has been observed in both CeCoIn$_5$ and PuCoIn$_5$\cite{kotroulakis13}.  This result is suggestive, but not conclusive. By contrast, observing the kink in $n_f(T)$ would provide conclusive evidence for composite pairing: in NpPd$_5$Al$_2$, the Np$^{4+}$ valence will increase smoothly with decreasing temperature as the $5f^3 \rarrow 5f^4$ fluctuations turn on, but then kink sharply downwards at $T_c$ as the $5f^3 \rarrow 5f^2$ fluctuations turn on (see Fig. 4(a)). This would be a 'smoking gun' signature of composite pairing.

\section{Conclusions}

While hidden order and superconductivity do not initially appear related, they can both be explained by two-channel Kondo physics, where the development of Kondo coherence breaks the symmetry of an excited doublet or pseudo-doublet at a phase transition, rather than the usual Kondo crossover. For hastatic order, this symmetry is time-reversal, while for composite pairing it is U(1) charge conjugation. The two phases have similar condensation entropies coming from their two-channel Kondo origin, $S \approx \frac{1}{2} R \log 2$ and similar magnetic field dependences, including quantum critical points at or near their upper critical fields, as magnetic field splits both ground state doublets.  Variations on the same non-Kramers theme should be relevant in other Pr and U compounds with different ground state doublets, as initially proposed in UBe$_{13}$ for the quadrupolar $\Gamma_3$\cite{ott83,cox88}. Another possible extension is to combine the hastatic and composite pair pictures to explain superconductivity in systems like Fig 1(b), but with the ground state and excited doublets switched, as seems likely in UBe$_{13}$\cite{ramirez}.

% \appendix

% \section{}
% \label{}

% The Acknowledgements are also a un-numbered section
\section*{Acknowledgements}
PC acknowledges support from NSF grant DMR-1309929.


\begin{thebibliography}{00}
% please try to use the bibitem system -
%   the references should be in order of citation in the text

\bibitem{mott}N.F. Mott, Phil. Mag. 30, 403 (1974).
\bibitem{doniach} S. Doniach, Physica B91, 231 (1977).
\bibitem{mathur08} N. D. Mathur et al, Nat. 394, 39 (1998); K. Miyake, S. Schmitt-Rink \& C. M. Varma, PRB 34, 6554 (1986); D. J. Scalapino, E. Loh\& J. E. Hirsch,PRB 34, 8190 (1986); M. T. Béal-Monod, C. Bourbonnais, \& V.J. Emery, PRB. 34, 7716 (1986).
\bibitem{extended} Q. Si, Physica B 378, 23 (2006); Phys. Status Solidi B 247, 476 (2010); Piers Coleman, Andriy H. Nevidomskyy, J. of Low Temp. Phys. 161, 182-202 (2010).
\bibitem{palstra85} T.T.M. Palstra  et al. Phys. Rev. Lett. 55, 2727 (1985). 
\bibitem{petrovic} C. Petrovic et al., J. Phys.: Condens. Matter {\bf 13}, L337(2001). % CeCoIn5
\bibitem{cecoir} C. Petrovic et al 2001 Europhys. Lett. 53 354 (2001).
\bibitem{aoki08} D. Aoki et al., J. Phys. Soc. Jpn., 76, 063701 (2008). 
\bibitem{coleman83} P. Coleman, Phys. Rev. B 28, 5255 (1983). 
\bibitem{newns}  N. Read and D.M. Newns, J. Phys. C {\bf 16}, 3273 (1983); ibid L1055, (1983).
\bibitem{chandra13} P. Chandra, P. Coleman and R. Flint, Nature 493, 621 (2013).
\bibitem{catk} P. Coleman, A. M. Tsvelik, N. Andrei and H. Y. Kee, Phys. Rev. B 60, 3608(1999). 
\bibitem{flint08} R. Flint, M. Dzero, and P. Coleman, Nature Physics 4, 643 (2008).
\bibitem{abrahams} E. Abrahams, A. Balatsky, D.J. Scalapino and  J. R. Schrieffer, Phys. Rev. B {\bf 52}, 1271(1995).
\bibitem{ceQCP} Johnpierre Paglione et al, Phys. Rev. Lett. 91, 246405 (2003).
\bibitem{npQCP} F. Honda et al J. Phys. Soc. Jpn. 77,  339 (2008).
\bibitem{uQCP} Y. S. Oh, Kee Hoon Kim, P. A. Sharma, N. Harrison, H. Amitsuka, and J. A. Mydosh, Phys. Rev. Lett. 98, 016401 (2007).
\bibitem{mydosh11} J. A. Mydosh and P. M. Oppeneer, Rev. Mod. Phys. 83, 1301 (2011).
\bibitem{broholm91} C. Broholm et al, Phys. Rev B 43, 12809 (1991).
\bibitem{park02} J.-G. Park, K. A. McEwen, M. J. Bull, Phys. Rev. B 66, 094502 (2002).
\bibitem{jeffries10} J.R. Jeffries et al., Phys. Rev. B 82, 033103 (2010).
\bibitem{altarawneh12} M. M. Altarawneh et al.  Phys. Rev. Lett. 108, 066407 (2012). 
\bibitem{amitsuka94} H. Amitsuka and T. Sakakibara, J. Phys. Soc. Japan 63, 736 (1994).
\bibitem{flint13} Rebecca Flint, Premala Chandra and Piers Coleman, {\bf Phys. Rev. B 86}, 155155 (2012).
\bibitem{schmidt10} A.R. Schmidt et al.Nature 465, 570–576 (2010). 
\bibitem{aynajian10} P. Aynajian et al, Proc. Natl Acad. Sci. USA 107, 10383–10388 (2010). 
\bibitem{dakovski11} G.L. Dakovski et al, Phys. Rev. B 84, 161103(R) (2011).
\bibitem{villaume08} A. Villaume, F. Bourdarot, E. Hassinger, S. Raymond, V. Taufour, D. Aoki, and J. Flouquet, Phys. Rev. B 78, 012504 (2008).
\bibitem{das13} P Das, R E Baumbach, K Huang, M B Maple, Y Zhao, J S Helton, J W Lynn, E D Bauer and M Janoschek, {\bf New J. Phys. 15}, 053031 (2013).
\bibitem{metoki13} N. Metoki, Hironori Sakai, Etsuji Yamamoto, Naoyuki Tateiwa, Tatsuma Matsuda, and Yoshinori Haga, {\bf J. Phys. Soc. Jpn. 82},055004  (2013).
\bibitem{ross14} K.A. Ross, L. Harriger, Z. Yamani, W. J. L. Buyers, J. D. Garrett, A. A. Menovsky, J. A. Mydosh, and C. L. Broholm, arXiv:1402.2689 (2014).
\bibitem{chandra14} Premala Chandra, Piers Coleman, Rebecca Flint, arXiv:1404.5920 (2014).
\bibitem{bernal} O.O. Bernal, M.E. Moroz, D.E. MacLaughlin, H.G. Lukefahr, J.A. Mydosh, T.J. Gortenmulder, %``Ambient Pressure $^{99}$Ru NMR in $URu_2Si_2$:  Internal Field Anisotropy,''
{\bf J. Mag. Magn. Mat. 272}, {E59-60} (2004).
\bibitem{caciuffo}  Private communication from R. Caciuffo.
\bibitem{okazaki11} R. Okazaki. et al. Science 331, 439–442 (2011).
\bibitem{matsuda} Private communication from Y. Matsuda
\bibitem{flint10} R. Flint and P. Coleman, Phys. Rev. Lett. 105, 246404 (2010).
\bibitem{sarrao07} J.L. Sarrao\& J. D. Thompson, J. Phys. Soc. Jap. 76, 051013(2007).
\bibitem{paglione07} J. P. Paglione et al, Nat. Phys. 3, 703 (2007); L. Shu et al, PRL. 106, 156403 (2011).
\bibitem{sn} E. D. Bauer et al Phys. Rev. B 73, 245109 (2006).
\bibitem{flint11} R. Flint, A. Nevidomskyy and P. Coleman, Phys. Rev. B 84, 064514 (2011).
\bibitem{gofryk09} K Gofryk et al, Phys. Rev. B 79, 134525 (2009).
\bibitem{kotroulakis13} G. Koutroulakis, H. Yasuoka, H. Chudo, P. H. Tobash, J. N. Mitchell, E. D. Bauer, J. D. Thompson, arXiv:1312.5404 (2013).
\bibitem{ott83} H. R. Ott et al, Phys. Rev. Lett. 50, 1595 (1983).
\bibitem{cox88} D.L. Cox, Physica C 15, 1642 (1988).
\bibitem{ramirez} A. P. Ramirez, P. Chandra, P. Coleman, Z. Fisk, J. L. Smith, and H. R. Ott, Phys. Rev. Lett. 73, 3018 (1994).

%\bibitem{bauer12} E. D. Bauer et al, J. Phys.: Condens. Matter 24 052206 (2012).
%\bibitem{nakatsuji08} S. Nakatsuji et al, Nature Physics4, 603 - 607 (2008).
%\bibitem{frings83} P.H. Frings et al. J. Magn. Magn. Mater. 31, 240 (1983).
%\bibitem{sarrao02} J. L. Sarrao et al., Nature (London) 420, 297 (2002). 
\end{thebibliography}
\end{document}